\documentclass[%
 reprint,
showpacs,preprintnumbers,
 amsmath,amssymb,
 aps,
]{revtex4-1}

\newcommand{\afs}{\rm $\alpha$-FeSi$_{2}$}

\usepackage{graphicx}
\usepackage{dcolumn}
\usepackage{bm}

\begin{document}

\preprint{APS/123-QED}

\title{Protected Fe valence in quasi-two dimensional $\alpha$-FeSi$_2$}

\author{W. Miiller$^{1}$, J. M. Tomczak$^{2,3}$, J.W.Simonson$^{1}$, G. Smith$^{1}$, G. Kotliar$^{2}$ and M. C. Aronson$^{1,4}$} 

\affiliation{$^{1}$Department of Physics and Astronomy, Stony Brook University, Stony Brook, New York 11794, USA}
\affiliation{$^{2}$Department of Physics and Astronomy, Rutgers University, Piscataway, New Jersey 08854, USA}
\affiliation{$^{3}$Institute of Solid State Physics, Vienna University of Technology, A-1040 Vienna, Austria}
\affiliation{$^{4}$Condensed Matter Physics and Materials Science Department, Brookhaven National Laboratory, Upton, New York 11793, USA}

\date{\today}

\begin{abstract}
We report the first comprehensive study of the high temperature form ($\alpha$-phase) of iron disilicide.
Measurements of the magnetic susceptibility, magnetization, heat capacity and resistivity were performed on 
well characterized single crystals.
With a nominal iron $d^6$ configuration, and a quasi-two dimensional crystal structure that strongly resembles that of LiFeAs,  $\alpha$-FeSi$_2$ is a potential candidate for unconventional superconductivity. 
Akin to LiFeAs, $\alpha$-FeSi$_2$ does not develop any magnetic order, and we confirm its metallic state down to the lowest temperatures ($T$=1.8 K).
However, our experiments reveal that paramagnetism and electronic correlation effects in $\alpha$-FeSi$_2$ are considerably weaker than in the pnictides.
Band theory calculations yield small Sommerfeld coefficients of the electronic specific heat $\gamma=C_e/T$ 
that are in excellent agreement with experiment.
Additionally, realistic many-body calculations further corroborate that quasi-particle mass enhancements are only modest in \afs\ .
 Remarkably, we find that the natural tendency to vacancy formation 
in the iron sublattice has little influence on the iron valence and the density of states at the Fermi level. Moreover, Mn doping does not significantly change the electronic state of the Fe ion. This suggests that the iron valence is 
protected against hole doping, and indeed the substitution of Co for Fe causes a rigid-band like response of the electronic properties. 
As a key difference from the pnictides, we identify 
the smaller inter-iron layer spacing, which causes the active orbitals near the Fermi level to be
of a different symmetry in $\alpha$-FeSi$_2$.
This change in orbital character might be responsible for the lack of 
superconductivity in this system, providing constraints
on pairing theories in the iron based pnictides and chalcogenides.
\end{abstract}

\maketitle

\section{Introduction}
\label{Introduction}

The newest class of 
unconventional
superconductors with transition temperatures as high as 55 K has stimulated the exploration for new Fe-based materials. Two main groups of materials may be distinguished:  systems like LaFeAsO and BaFe$_{2}$As$_{2}$\cite{RevModPhys.83.1589}, where superconductivity appears in the vicinity of the disappearance of magnetic order and is tunable by doping or external pressure, and stoichiometric systems like LiFeAs \cite{PhysRevB.78.060505} where  superconductivity is the ground state. The latter compound has a quasi-two dimensional crystal structure, where square nets of Fe atoms are tetrahedrally coordinated with As ions, while Li ions are placed between the Fe-As sheets.  There is no evidence for magnetic 
order in LiFeAs, although pronounced antiferromagnetic correlations were observed in the normal state \cite{PhysRevB.83.220514}, where the electronic configuration of iron ions is close to $d^6$. For a long time only Fe-based compounds with pnictogens or chalcogenides have been found to reveal high-temperature superconductivity, but the recent finding of superconductivity in YFe$_2$Ge$_2$, believed to be in the proximity of an antiferromagnetic quantum critical point \cite{zou2013fermi}, increases the interest in tetralide containing intermetallics, based on the elements Si, Ge, Sn, and Pb.

In this report we focus on just such a system, \afs\ , which is very similar to LiFeAs from the structural and electronic point of view. \afs\ is the most iron deficient phase in the Fe-Si binary phase diagram \cite{massalski2001binary}, existing in two different allotropes. The orthorhombic form, $\beta$-FeSi$_2$, is stable at room temperature and was characterized as a wide gap semiconductor \cite{PSSB:PSSB19690340269}. In contrast, the much simpler tetragonal structure, $\alpha$-FeSi$_2$, is metallic and only stable above 965$^o$C. \afs\ can, however, be studied at room temperature and below by quenching the material from high temperatures, and a few crystal structure \cite{dubrovskaya1962structure,gueneau1995quantitative}, magnetic susceptibility \cite{PSSB:PSSB19690340270}, heat capacity \cite{Acker19991523}, and M\"{o}ssbauer spectroscopy \cite{0022-3719-6-14-017,0953-8984-13-11-102} measurements carried out on a variety of single crystal and polycrystalline samples exist. However no detailed single-crystal study has been performed to date. In view of the metastable nature of \afs, it is crucial to obtain a complete suite of experiments carried out on single crystals where the structure and morphology are well explored and reproducible. This is the aim of the present work.\\
Our results confirm that \afs\ is an excellent metal with minimal electronic correlations, consistent with a nonmagnetic $d^{6}$ configuration for the Fe ions. A complete investigation of the crystal structure has been carried out using single crystal X-ray diffraction, finding that there are substantial numbers of Fe vacancies, reflected in the obtained composition Fe$_{0.83}$Si$_2$. A wide range of Fe deficiency was previously observed for this compound \cite{aronsson1960note}.
Electronic structure calculations were carried out 
for both stoichiometric \afs\ and an iron deficient supercell Fe$_{0.875}$Si$_2$, which is close to the actual composition. 
We find that the density of states at the Fermi level, $N(E_{F})$, is essentially the same in both cases.
Moreover, the 
theoretical
Sommerfeld coefficient 
is in excellent agreement with the measured results obtained from electronic heat capacity, $\gamma=C_e/T$. Dynamical mean field theory (DMFT) \cite{rmp_dmft} calculations corroborate that correlation effects do not play a significant role in this material.
Similarly, we find 
experimental evidence for the robustness of the electronic properties against hole doping, whereas introduction of electrons increases strongly both the magnetic susceptibility and the Sommerfeld coefficient, providing evidence of an increase in the number of states 
at the Fermi level.

\section{Experimental and computational details}
\label{Experimental}

Single crystals of pure and doped \afs\ were prepared using a Ga flux, and flat plate-like crystals with approximate dimensions of $5 \times 5 \times 0.1$ mm were obtained by quenching the melt from 980 $^{o}$C. The crystal structure  was determined at room temperature using crystals with approximate dimensions of $0.05\times 0.2\times 0.2$ mm$^3$ using a Bruker Apex-II single crystal diffractometer with Mo-K$\alpha$ radiation. Refinements of these data were carried out using the programs Jana and Superflip~\cite{petvrivcek2006jana2006}.

Measurements of the magnetization $m$ were carried out in temperatures $T$ ranging from 1.8 to 400 K and in magnetic fields $B$ as large as 7 T using a Quantum Design Magnetic Properties Measurement System (MPMS) on a co-aligned collection of approximately 20 mg of crystals that were wrapped in gold foil. Electrical contacts were made to the crystal using silver-filled epoxy in the four-probe configuration, and electrical resistivity $\rho$ measurements were performed for temperatures from 1.8 - 300 K in a Quantum Designs Physical Properties Measurement System (PPMS), where the 1 mA current flowed in the \emph{a-b} plane. Heat capacity $C_p$ measurements were also carried out from 1.8 -300 K using a PPMS.

Using our experimental atomic positions, 
electronic structure calculations were performed within the local density approximation (LDA) as implemented in wien2k\cite{Blaha1990399} for both 
stoichiometric $\alpha$-FeSi$_2$ and the supercell Fe$_{7}$Si$_{16}$, which is close to the actual composition Fe$_{0.83}$Si$_{2}$.
Realistic many-body calculations were performed in the  LDA+DMFT framework of Ref.~\onlinecite{PhysRevB.81.195107}.
For the Hubbard interaction and the Hund's rule coupling, we employed the values $U=5$eV and $J=0.7$eV that were proven to be reasonable
for Fe-pnictides\cite{2011NatPh...7..294Y,PhysRevB.82.045105} and other iron silicides\cite{jmt_fesi}.
To allow for a direct comparison with the pnictide LiFeAs, all theoretical calculations of \afs\ were done in a $\sqrt{2}\times\sqrt{2}$
non-primitive cell with 2 Fe atoms.

\section{Results and discussion}
\label{Results and discussion}

\begin{figure} [h]
\includegraphics[width=9cm]{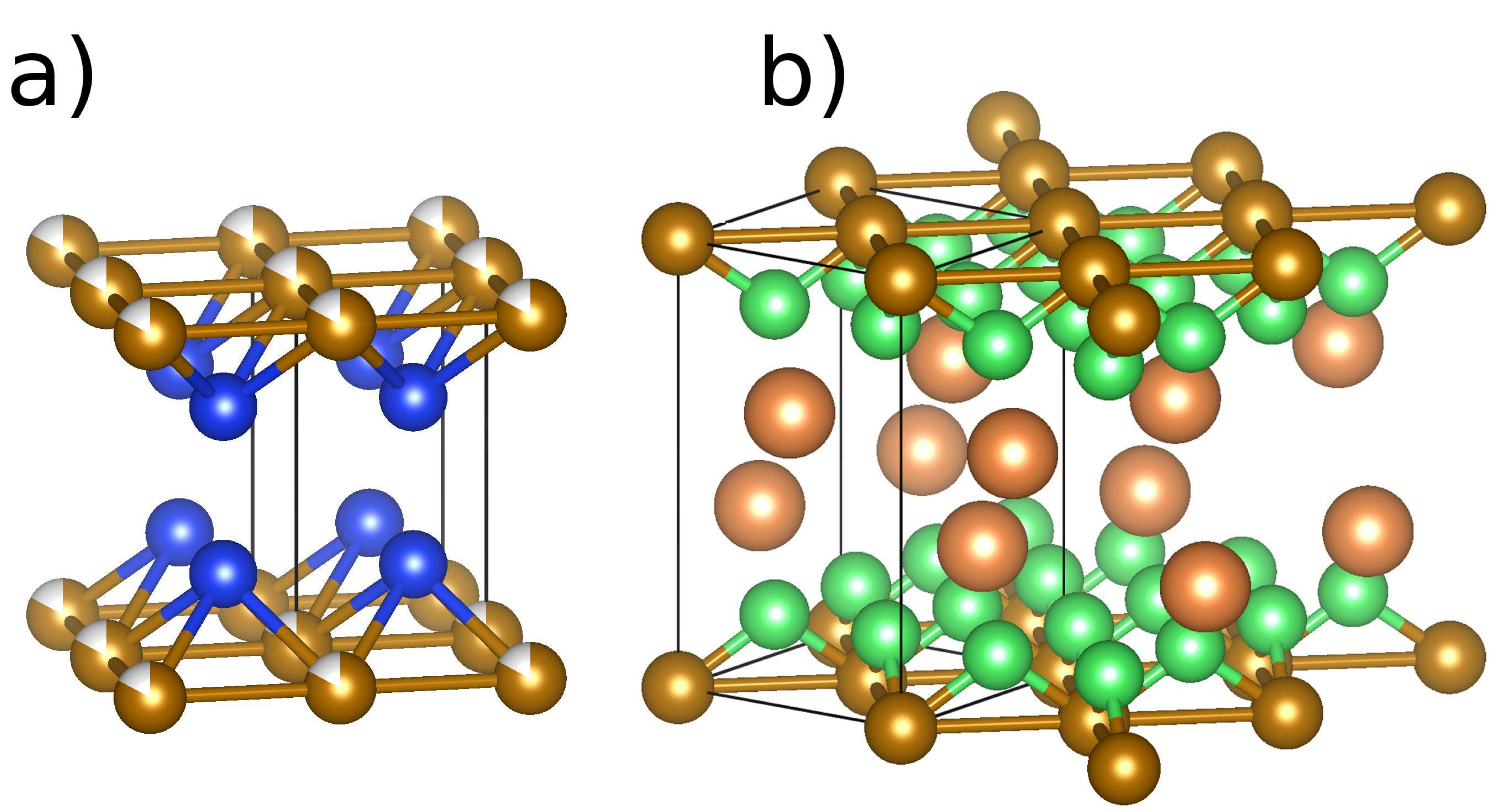}%
\caption{\label{structure}(Color online) Crystal structure of $\alpha-$FeSi$_2$(a) compared to LiFeAs(b). Brown, blue, orange and green spheres represent Fe, Si, Li and As ions, respectively. Unit cells are indicated.}
\end{figure}

Energy Dispersive X-ray Analysis (EDX) measurements of the single crystals revealed that the chemical composition is  Fe$_{0.83(1)}$Si$_2$. No impurities or contaminant phases were detected. The results of single crystal X-ray diffraction measurements are shown in Table \ref{table1}. The centrosymmetric space group $P4/mmm$ was confirmed with lattice parameters $a=b$=2.6955 \AA{} and $c=5.1444$ \AA{}, in  good agreement with previously reported values obtained from polycrystalline samples \cite{dubrovskaya1962structure,gueneau1995quantitative}. A Rietveld refinement of the single crystal X-ray data yielded the composition Fe$_{0.832}$Si$_2$, consistent with EDX results. The crystal structure of \afs\ is compared to that of LiFeAs in Fig. \ref{structure}. It consists of square Fe-deficient planes with the Fe-Fe spacing $d_{Fe-Fe}=a=2.6955$ \AA{} and interplanar distance $c=5.1444$ \AA{}. Both $d_{Fe-Fe}$ and $c$ are similar to values found in the superconducting iron pnictides, and especially in LiFeAs, where $d_{Fe-Fe}=a=2.6809$ \AA{} and $c=6.3639$ \AA{} (Fig. \ref{structure}b)\cite{PhysRevB.78.060505}. The Fe-Si distance, $d_{Fe-Si}=$2.35 \AA{} and is comparable to the equivalent quantity in LiFeAs, $d_{Fe-As}=$2.416 \AA{} whereas both ligand-Fe-ligand angles $\Phi(FeSi_2)=107.76^o$ and $\Phi(LiFeAs)=102.793^o$ are lower than the optimal value of 109.5$^o$\cite{JPSJ.77.083704}. The smaller interlayer separation in $\alpha$-FeSi$_2$ may result from the lack of alkali metal ions separating the $p$-metalloid layers.
\begin{center}
\begin{table} [h]
 \begin{tabular}{c| c| c| c| c| c| c| c|}
 \ & $x$ & $y$  & $z$ & Occ.* & 1000$U_{11,}$ & 1000$U_{33}$ ($\AA^2$) & 1000$U_{izo}$ ($\AA^2$) \\
  & &   &  &  & 1000$U_{22}$ ($\AA^2$) &  &  \\
 \hline
 Fe & 0 & 0 & 0 & 0.832 & 3.7 & 5.1 & 4.17 \\
 Si & 0.5 & 0.5 & 0.272 & 1 & 8.2 & 4.8 & 7.1\\
 \end{tabular}
 \caption{\label{table1}Structural parameters of $\alpha$-FeSi$_2$ determined from refinement of X-ray Laue diffraction data collected at room temperature. $R=1.28$, $R_w=2.19$, GOF$=1.94$ for 986 reflections with $I\geq 3\sigma (I)$. *Fractional occupancy.}
 \end{table}
\end{center}

We have also grown doped $\alpha$-Fe$_xT_y$Si$_2$ crystals, where $T=$ Mn and Co.  EDX analysis, as well as single crystal X-ray diffraction, confirmed the successful introduction of dopants into the structure. We have obtained samples with Mn$_{0.08}$Fe$_{0.74}$Si$_2$, Mn$_{0.04}$Fe$_{0.74}$Si$_2$, Co$_{0.05}$Fe$_{0.7}$Si$_2$ and Co$_{0.1}$Fe$_{0.7}$Si$_2$ compositions, according to EDX. Unfortunately, as Mn, Fe and Co are almost isoelectronic, exact determination of compositions from single crystal diffraction was impossible. All compositions reported here were carefully determined by the EDX technique.

\begin{figure} [h]
\includegraphics[width=9cm]{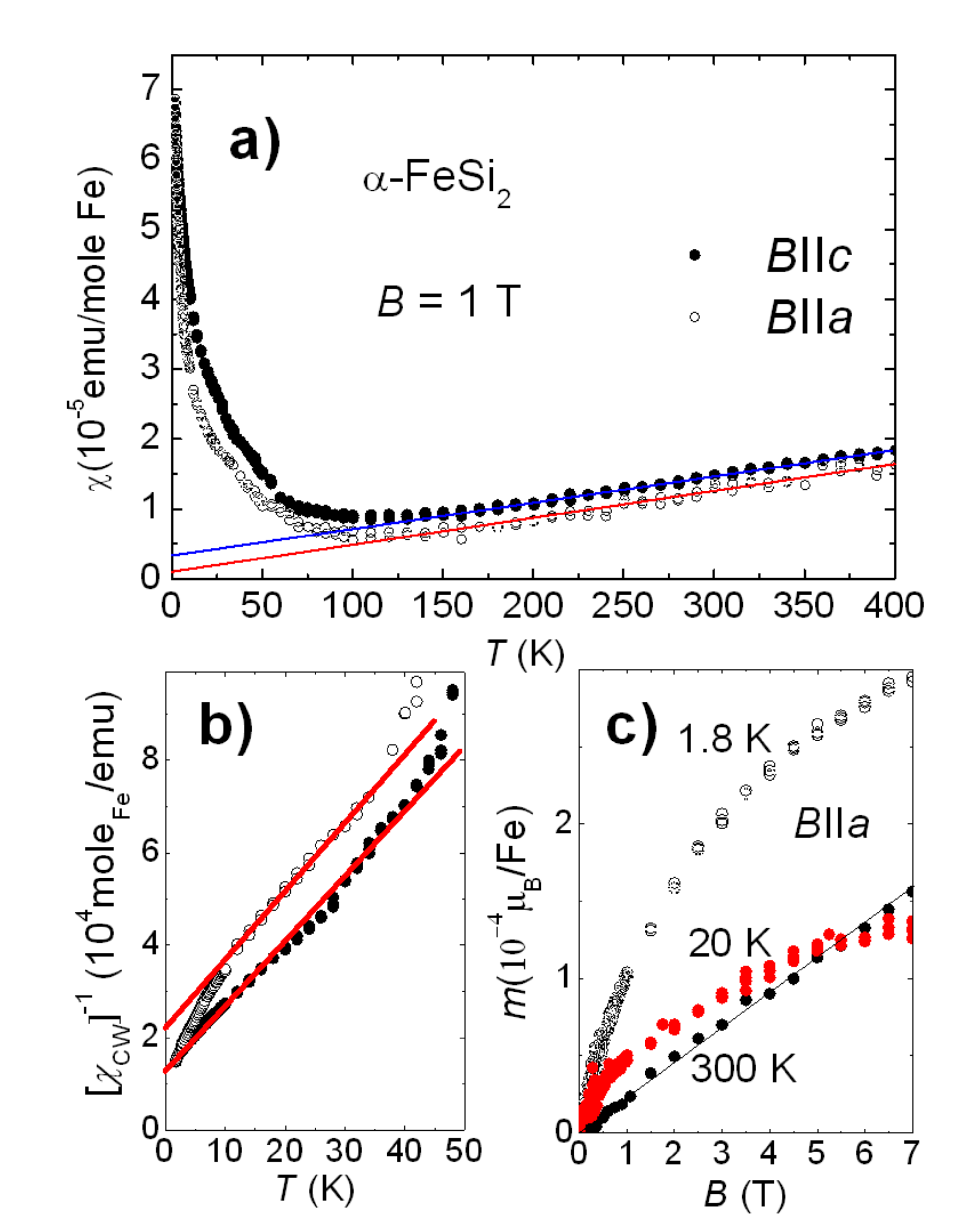}%
\caption{\label{sus_fig}(Color online) a) Magnetic susceptibility of $\alpha-$FeSi$_2$ and magnetization collected for several temperatures along the principal axes. For details see the text. b) Curie-Weiss fit performed below 50 K on the $\chi_{CW}$ extracted from the experimental data, c) Magnetization collected at 1.8, 20 and 300 K.}
\end{figure}

The temperature dependencies of the molar susceptibilities, $\chi=m/H$, of $\alpha$-FeSi$_2$, measured with a 1 T field along the $a$ and $c$ axes, are shown in Fig. \ref{sus_fig}a. Both curves reveal a weak temperature dependence with Curie-Weiss like tails developing below $\simeq$ 100 K and linear increases in $\chi(T)$ above this temperature. In particular, there is no suggestion of magnetic order below room temperature.  These results are qualitatively consistent with data previously reported on polycrystalline samples\cite{PSSB:PSSB19690340270}. We have extracted and fitted the low-temperature paramagnetic tail, $\chi_{CW}(T)$ to a Curie-Weiss expression (the fit is shown in Fig. \ref{sus_fig}b), and the magnitude of the effective moment 7.5$\times10^{-2}\mu_B/$Fe amounts to no more than $\simeq$  0.2\% of Fe$^{3+}$ impurities in the low-spin ($S=1/2$) state per mole of $\alpha$-FeSi$_2$. It is likely that these moments are associated with paramagnetic impurities, and that the low temperature tail is extrinsic in origin. The magnetization is plotted as a function of the applied magnetic field at 1.8, 20 and 300 K in Fig. \ref{sus_fig}c. At the lowest temperatures, the magnetization of $\alpha$-FeSi$_2$ seems to be dominated by a nonlinear component that we have attributed to the paramagnetic impurities, as the intrinsic response is expected to be linear in field and small.

$\chi$ is almost isotropic, and  is in the range 10$^{-6}$ - 10$^{-5}$ emu/mole$_{Fe}$. This is $\simeq$100 times smaller than the values found in the Fe-pnictdes and chalcogenides\cite{RevModPhys.83.1589}, but close to the values found in  simple, paramagnetic metals.  Fig. ~\ref{sus_fig}a shows a linear increase of the magnetic susceptibility with increasing temperature above $T=100$ K. This feature was commonly observed among iron pnictides and has been associated with antiferromagnetic spin fluctuations \cite{1367-2630-11-4-045003}. However. it is also predicted theoretically for metallic systems where the  Fermi level ($E_F$) is located in a dip in the density of states N(E$_{F}$) \cite{PhysRev.94.837}. Indeed, in a number of transition metals (Cr, Mo, Ti \cite{1961RSPSA.260..237K}) and paramagnetic intermetallics (e.g. CoSi \cite{PhysRevB.86.064433}) a linear increase in the temperature dependence of the magnetic susceptibility was also observed, although at very low temperatures an increase of $\chi(T)$ with decreasing temperature is predicted \cite{shimizu1960magnetic}. As we will describe below, the band structure calculations find such a minimum in the density of states, indicating that the latter mechanism is the likely origin of the increase in $\chi$(T) observed in $\alpha$-FeSi$_{2}$.

Since the Curie-Weiss contribution from magnetic impurities is negligible above 100 K, the susceptibility data from that region are fitted in Fig.~\ref{sus_fig} to the expression $\chi(T) = \chi(0)+A\times T$, where  $\chi(0)$ is the susceptibility in the zero-temperature limit, with the average value of $\chi(0)=3.4\times10^{-6}$ emu/mole-Fe, along both $a$ and $c$ axes.
 $\chi(0)$ consists of two contributions: a Pauli paramagnetic part $\chi(0)_p$, related to the density of states at the Fermi level, and $\chi(0)_{dia}$, arising from the diamagnetic response of the Si and Fe core electrons. Taking $\chi(0)_{dia}(Si^{4+})= -3.9\times10^{-6}$ emu mole$^{-1}$ \cite{PhysRevA.2.1130} and $\chi(0)_{dia}(Fe^{2+})= -19.2\times10^{-6}$ emu mole$^{-1}$ \cite{9120}, yields $\chi(0)_p \simeq$ 2.5$\times 10^{-5}$emu/mole-Fe.

\begin{figure} [t]
\includegraphics[width=9cm]{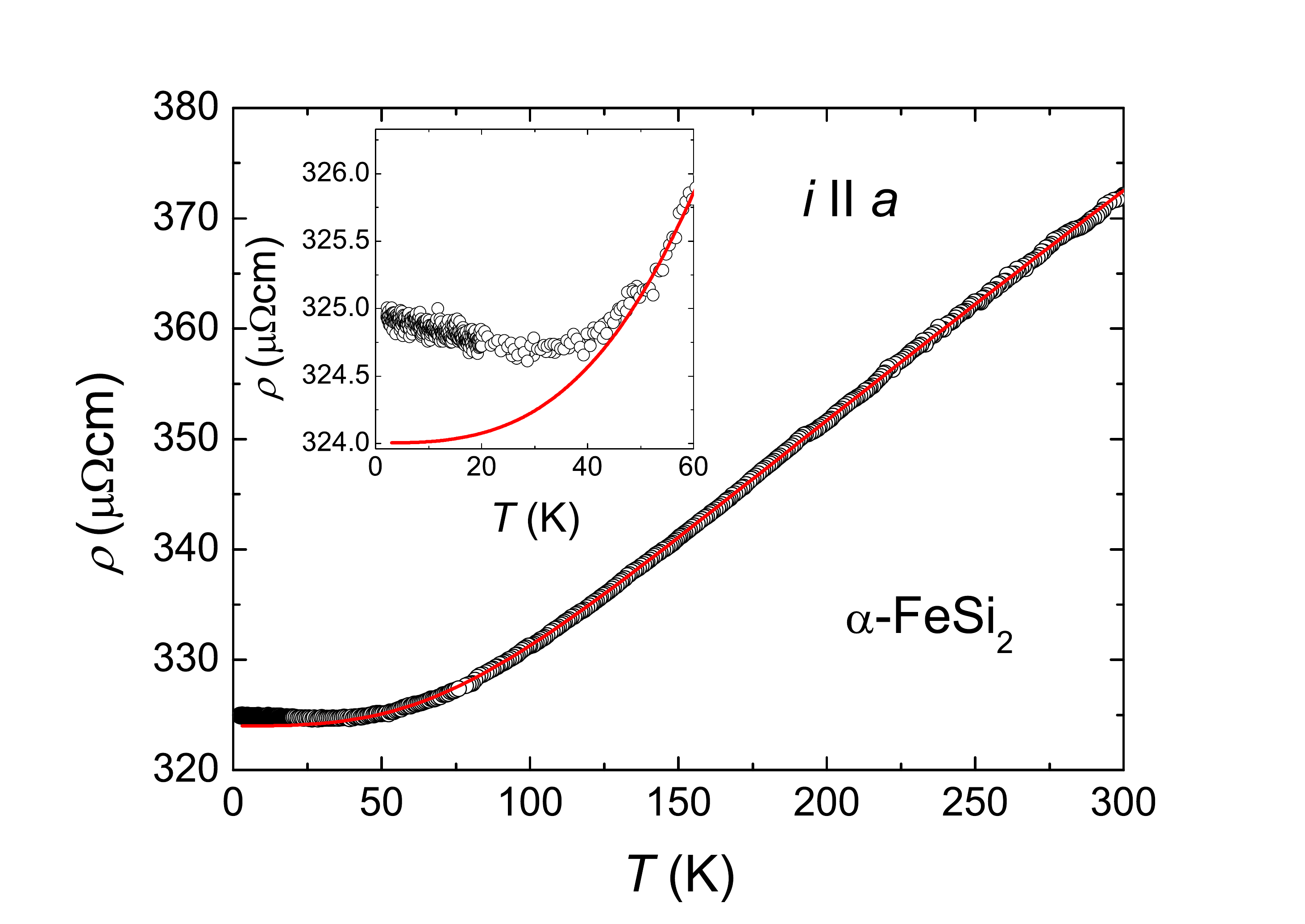}%
\caption{\label{fig_res}(Color online) Resistivity ($\rho$) measured along \textit{a} axis as a function of temperature together with a fit to eq. \ref{resistivity-formula}. The inset shows an enlargement of the low-temperature data.}
\end{figure}

The temperature variation of the electrical resistivity $\rho$(T), measured with the current flowing along the $a$ axis, is shown in Fig. \ref{fig_res}. The resistivity $\rho$ decreases linearly with temperature from 372 $\mu \Omega$cm at 300 K to 325$\mu \Omega$cm at 40 K, in manner consistent with the Bloch-Gr\"{u}neisen expression:

\begin{equation}
\rho = \rho_0+4R\Theta_{D}^R(\frac{T}{\Theta_{D}^R})^5\int_0^{\Theta_D^R/T}\frac{x^5 \mathrm{d} x}{(e^x-1)(1-e^{-x})}
\label{resistivity-formula}
\end{equation}

Here, the residual resistivity $\rho_0=324$ $\mu \Omega$cm and the Debye temperature $\Theta_D^R$ =654 K. While $\rho$(T) is definitively metallic, the low residual resistivity ratio $\rho(300 K)/ \rho_0=1.14$ and the somewhat high value of $\rho_{0}$ may result  from the presence of vacancies in the Fe-square net plane. Below $\simeq$30 K, $\rho$(T) starts to increase (inset, Fig. \ref{fig_res}), suggesting incipient charge localization.

\begin{figure} [b]
\includegraphics[width=9cm]{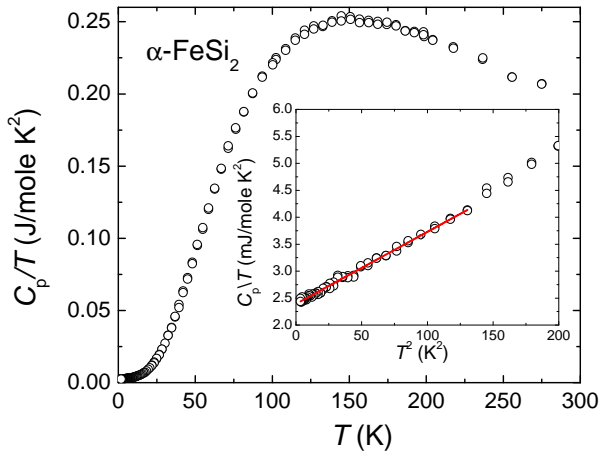}%
\caption{\label{Cp_1_fig}(Color online) Heat capacity over temperature ratio $C_p/T$, measured for $\alpha$-FeSi$_2$. The inset shows fit to the eq. 
\ref{eq-cpT} (for details see the text).}
\end{figure}

The temperature dependence of the molar heat capacity $C_p/T$ is depicted in Fig. \ref{Cp_1_fig}. No anomaly is observed for temperatures as low as 1.8 K, excluding any magnetic or structural transition. Below 15 K, $C_p/T$ is well described by the expression 
\begin{equation}
C_p/T = \gamma+\beta T^2
\label{eq-cpT}
\end{equation}

where the Sommerfeld coefficient, $\gamma$= 2.0(1) mJ/ mole-Fe K$^{-2}$, accounts for the electronic contribution, and the second term represents the contribution of the phonons to $C_{P}$, with $\beta$=$13.4\times 10^{-3}$ mJ/mole K$^{4}$ (see the inset in Fig. \ref{Cp_1_fig}). We use the Debye model to calculate the Debye temperature $\theta_{D}$= 744 K from $\beta$.
 
\begin{figure}
\includegraphics[width=9cm]{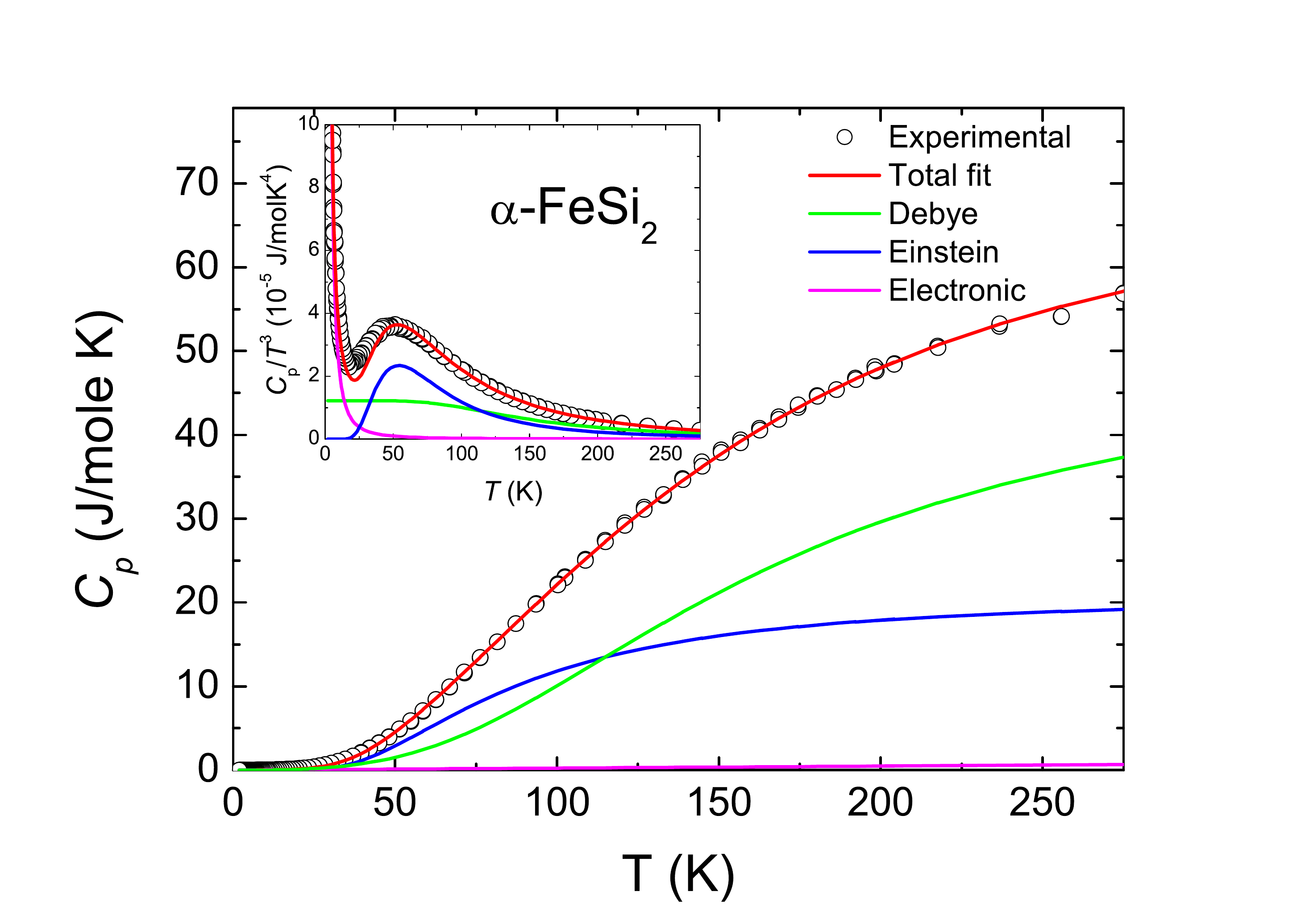}%
\caption{\label{fig1}(Color online)Deconvolution of the total heat capacity $C_p$ into electronic and lattice terms. The inset shows maximum in $C_p/T^3(T)$ dependence, being a significant hallmark of Einstein modes. Solid lines represent lattice (Debye and Einstein) and electronic contributions to the heat capacity, according to the eq. \ref{eq:9}.}
\end{figure}

As the total heat capacity of solids in general consists of lattice, magnetic and electronic contributions, one can decompose 
$C_p$ into different components. In order to investigate the vibrational properties of Si and Fe sublattices, we have modelled the phonon contribution to the heat capacity by different combinations of Debye and Einstein modes\cite{PhysRevB.27.1568}. The best agreement with experimental data is shown in Fig. \ref{fig1}, coming from a model that assumes Si-based Debye modes that have 2 atoms per unit cell $n_{D}$=2, and Einstein modes, attributed to the $n_{E}$=0.83 Fe atoms per unit cell. The overall $C_p(T)$ may then be well described with the formula:

\begin{equation}
\begin{array}{rcl}
C_p& =& \gamma T+9Rn_D\int_0^{\Theta_D/T}\frac{x^4e^x}{(e^x-1)^2}dx+\\ & & 3Rn_E(\frac{\Theta_E} {T})^2\frac{e^{\frac{\Theta_E}{T}}}{(e^{\frac{\Theta_E}{T}}-1)^2}
\end{array}
\label{eq:9}
\end{equation}

Here, $R$ is the gas constant, $\Theta_D$=683 K is the Debye temperature, and $\Theta_E$=266 K is the Einstein temperature. We note that the values of $\Theta_D$ that were determined from heat capacity and electrical resistivity data are comparable. The inset of Fig.~5 shows the quantity $C_{\rm{p}}/T^3$, to highlight that the maximum at $\simeq$ 50 K is well reproduced by the Einstein modes represented in our fit. Similar Einstein-like modes were also inferred from fits to the heat capacity in several  Fe-based superconductors~\cite{1367-2630-11-2-025010}.

The small value of $\gamma$ points to a low density of states at the Fermi level. We calculate the Wilson ratio 
$R_{W}=k_{B}^{2}\pi^{2}\chi(0)_{P}/(3\mu_{B}^{2}\gamma)$ =0.9, which is very close to the value R$_{W}$=1 that is expected for a Fermi liquid with no mass enhancement. The picture that emerges from our susceptibility, resistivity, and heat capacity measurements is that $\alpha$-FeSi$_{2}$ is a very weakly correlated metal with weak itinerant paramagnetism. This suggests that \afs\ may be well described by \textit{ab initio} effective one-particle band-structure methods. 

\begin{figure}
\includegraphics[width=9cm]{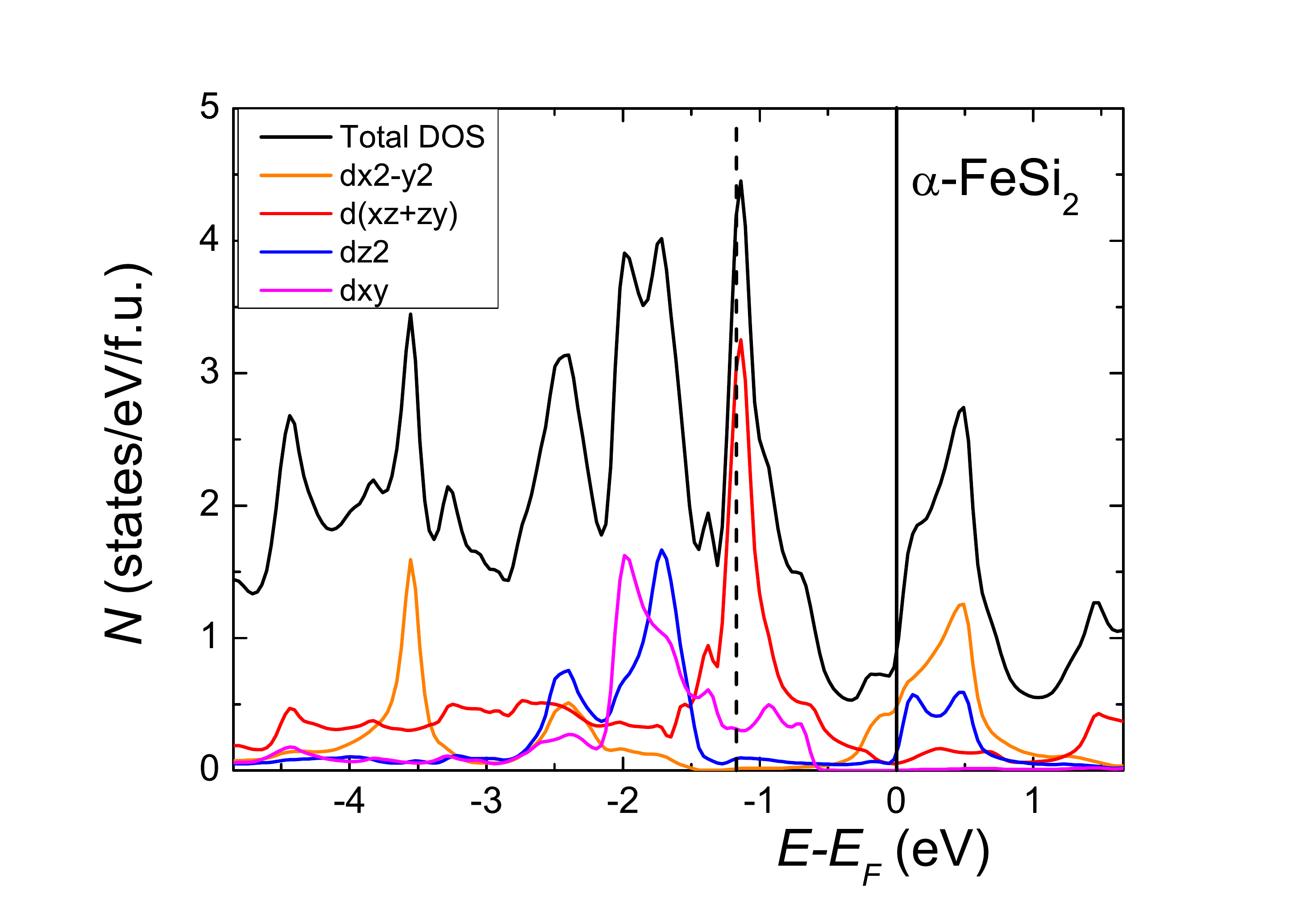}%
\caption{\label{DOS}(Color online) LDA density of states, $N(E)$, calculated for stoichiometric $\alpha$-FeSi$_2$. Solid line indicates Fermi level yielded by the calculations, whereas the dashed one represents the Fermi level estimated from a rigid-band approach.}
\label{LDA}
\end{figure}

We begin the electronic structure calculations with the stoichiometric $\alpha$-FeSi$_2$ composition using the density functional formalism within LDA.
The resulting density of states (DOS) displayed in Fig.~\ref{LDA} is in good agreement with previous calculations \cite{PhysRevB.59.12860}.
The density of the 3$d$ electrons of iron extend with shallow tails from -10 to beyond +10eV, leading to a very large bandwidth that signals a degree of charge carrier delocalization
that is notably larger than in the pnictide LiFeAs\cite{PhysRevB.78.094511}.
%
The electron count of the Fe 3$d$ states in \afs\ is 6.7 (within LDA, by orbital projection \cite{PhysRevB.81.195107}).
%
%
 The total density of states at the Fermi level $N(E_F)$ is 0.9 states eV$^{-1}$ per Fe, which is substantially smaller than that found for the superconductor LiFeAs 
 (1.93 states/eV per Fe)\cite{2008JETPL..88..543N}, 
  LaOFeAs (2.0) and BaFe$_2$As$_2$ (2.11)\cite{2008JETPL..88..543N}. 
  Fig \ref{DOS} shows that  the states at E$_{F}$ in $\alpha$-FeSi$_{2}$ are primarily derived from the $d_{x^2-y^2}$ orbital.\footnote{The orbital characters refer to the $\sqrt{2}\times\sqrt{2}$ non-primitive cell with 2 Fe atoms and space group 129, in which the iron planes are oriented as in the conventional cell of LiFeAs.} 
Since for the iron pnictides it has been suggested that the presence of the $d_{xy}$, $d_{yz}$ and $d_{zx}$ orbitals at the Fermi level play an active role in the superconductivity \cite{nekrasov2008electronic}, the decided lack of these orbitals at the Fermi level might be at the heart of the absence of superconductivity in $\alpha$-FeSi$_2$.
  
 The orbital character of the bands at the Fermi level is the result of the smaller interlayer spacing in $\alpha$-FeSi$_{2}$. In LiFeAs, the iron/pnictogen layers are separated by alkali metal ions, while the iron/silicon layers in $\alpha$-FeSi$_{2}$ hybridize 
  more strongly, which is also responsible for the very broad valence band. The calculated low density of states at the Fermi level suggests that both, superconducting or magnetic ground states are unfavourable. Indeed, our LSDA calculations that assume the stripe-like antiferromagnetic order common to the 122-pnictides%
\footnote{stripe-like anti-ferromagnetic (AF) order in-plane, and AF inter-layer order.},
find the staggered moment of \afs\ to be zero, in agreement with the absence of magnetic order.
 
The LDA calculations for stoichiometric $\alpha$-FeSi$_{2}$ reproduce several of our experimental observations. The Sommerfeld coefficient is calculated to be  $\gamma_{th}=\frac{\pi^2}{3}k_B^2N_AN(E_F)=2.1$ mJmol$^{-1}_{Fe}$K$^{-2}$, in  good agreement with the experimental value obtained for our Fe deficient crystal (2.9 mJmol$^{-1}_{Fe}$K$^{-2}$).  As noted above, the observed increase in $\chi(T)$ with increasing temperature (Fig.~\ref{sus_fig}) could, in principle, be explained if the Fermi level E$_{F}$ is located in or near a minimum of the density of states\cite{PhysRev.94.837}, just as we observe in \afs\ (Fig. 6).

The success of the LDA calculations is surprising, considering that the actual composition of our sample is Fe$_{0.83}$Si$_{2}$. If we assume that the deviation from stoichiometry results only in a shift of the Fermi level that leaves the underlying band structure unaffected, this iron deficiency corresponds to removing 17 $\%$ of the Fe-3d states. This  would shift (dotted line, Fig.~6) E$_{F}$ 1.17 eV below the value found for stoichiometric $\alpha$-FeSi$_{2}$ (solid line, Fig.~6). The new value of N(E$_{F}$) for Fe$_{0.83}$Si$_2$ becomes 4.4 states/eV \textit{per} formula unit, implying a value for $\gamma$=10.4 mJ/mole-Fe K$^{2}$ that is far in excess of the experimental value. Significantly, this new Fermi level requires a $d$-electron count of 5.3 electrons per iron, despite the essentially nonmagnetic character revealed in our susceptibility measurements. We conclude that this rigid band approximation fails in $\alpha$-FeSi$_{2}$.

\begin{figure} [h]
\includegraphics[width=9cm]{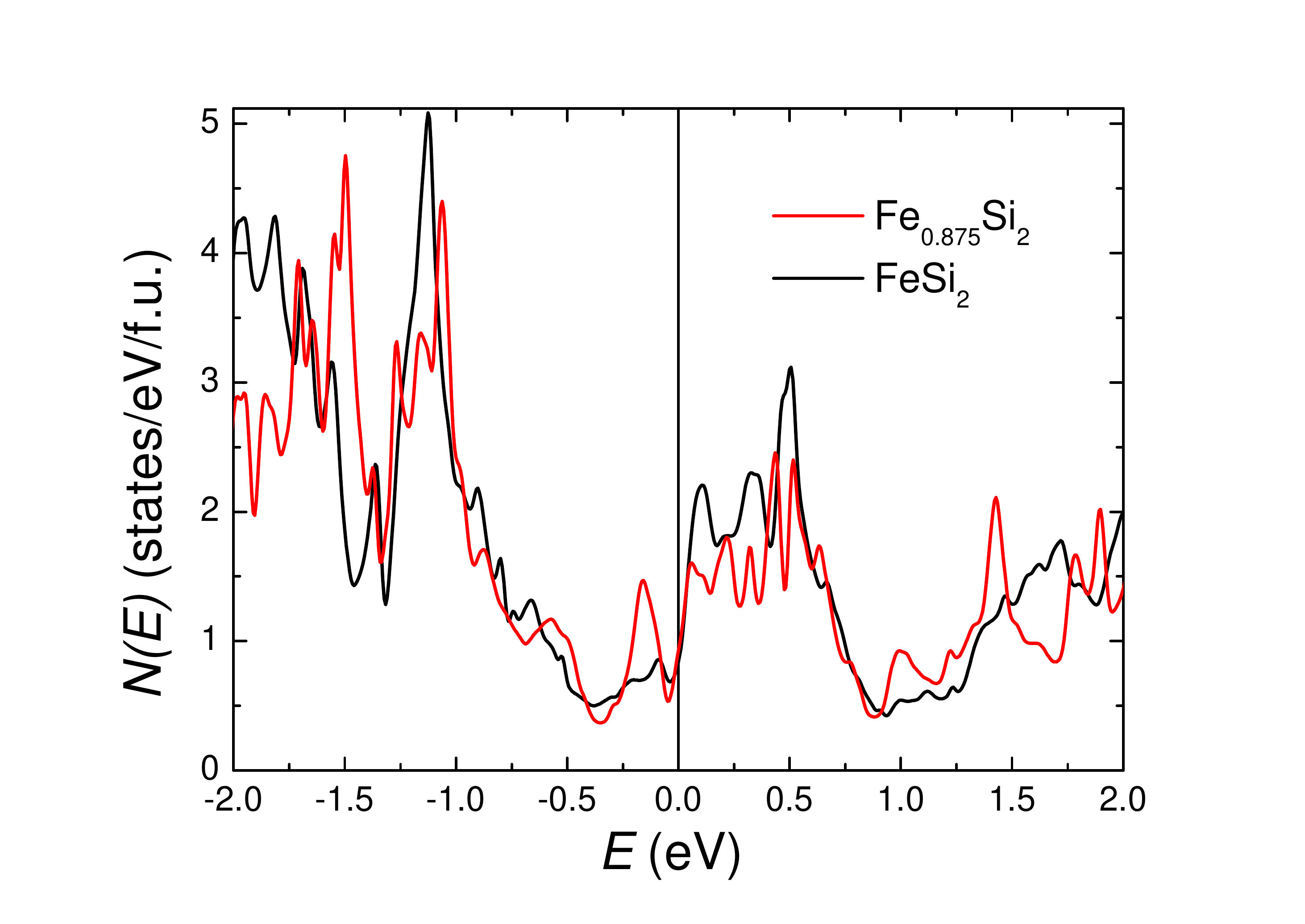}%
\caption{\label{Dos_SC_fig}(Color online) Total density of states calculated for 2x2x2 supercells with all Fe sites occupied (black line) and with one Fe vacancy introduced (red line). Vertical solid line shows Fermi level $E_F$.}
\end{figure}

To examine the electronic structure of \afs\ with a realistic Fe deficiency, we have performed calculations on a Fe$_7$Si$_{16}$ supercell, corresponding to the composition Fe$_{0.875}$Si$_2$.  
The resulting total DOS is displayed in Fig. \ref{Dos_SC_fig}. 
 Interestingly, the introduced vacancies do not alter the overall $N(E)$ significantly, especially in the proximity of the Fermi level. $N(E_F)$ is in both cases almost the same, as is the electronic occupation of the 3$d$ states per iron. The theoretical value of the Sommerfeld coefficient that results from these calculations for Fe$_{0.875}$Si$_2$, $\gamma_{th}=2.5$ mJ/mole$_{Fe}$ K$^{2}$ is in very good agreement with the experimental value $\gamma=2.9$ mJ/mole$_{Fe}$ K$^{2}$.
Moreover, the change in the (partial) charges of the iron and silicon atoms is found to be minimal, confirming the break down of the rigid-band approximation.
This preservation of the $d^6$ valence is also consistent with previous M\"{o}ssbauer measurements on nonstoichiometric $\alpha$-FeSi$_{2}$\cite{0022-3719-6-14-017}.

With this motivation, we performed additional LDA+DMFT calculations for stoichiometric \afs.
Correlation effects are found to be small.
The number of $d$-electrons is reduce slightly to 6.4 per iron, bringing the iron atoms closer to the nonmagnetic Fe$^{2+}$ state.
The many-body effective masses, resolved into orbital characters, are shown in Table \ref{table2}.
The mass enhancements are rather small, concurring that $\alpha$-FeSi$_2$ is only weakly correlated.
Indeed, values for LiFeAs are twice as large\cite{2011NatPh...7..294Y} but the mono-silicide FeSi exhibits
mass enhancements that are greater by at least 20\% \cite{jmt_fesi}. According to Yin  \textit{et al.} \cite{2011NatPh...7..294Y}, there is a correlation between effective band mass $m^*$ and such quantities as Fe-Pn(pnictogen) distance and Pn-Fe-Pn angle in the superconducting pnictides. According to these criteria, \afs\ should be much more correlated, with band mass enhancements in order of these observed in LiFeAs. The reason why such enhancement does not take place in \afs\ is its smaller interplanar Fe-Fe distance - 5.14 \AA , whereas typical plane separation in Fe-HTSC is in the range of 6-10 \AA. The presence of electronic correlations is then directly related to the more two-dimensional character of the pnictides.

\begin{table} [h]
\vspace{0.5cm}
 \begin{tabular}{c| c| c| c| c|}
 $m^{DMFT}/m^{LDA}$& $z^2$ & $x^2-y^2$  & $xz/yz$ & $xy$  \\
 \hline
 $\alpha$-FeSi$_2$ & 1.2 & 1.3 & 1.2 & 1.2 \\
LiFeAs\cite{2011NatPh...7..294Y} & 2.1 & 2.1 & 2.8 & 3.3 \\
  \end{tabular}
 \caption{\label{table2}Effective masses $m^{DMFT}/m^{LDA}$ for $\alpha$-FeSi$_2$ (at 290K) and LiFeAs\cite{2011NatPh...7..294Y}.}
 \end{table}

\begin{figure} 
\includegraphics[width=9cm]{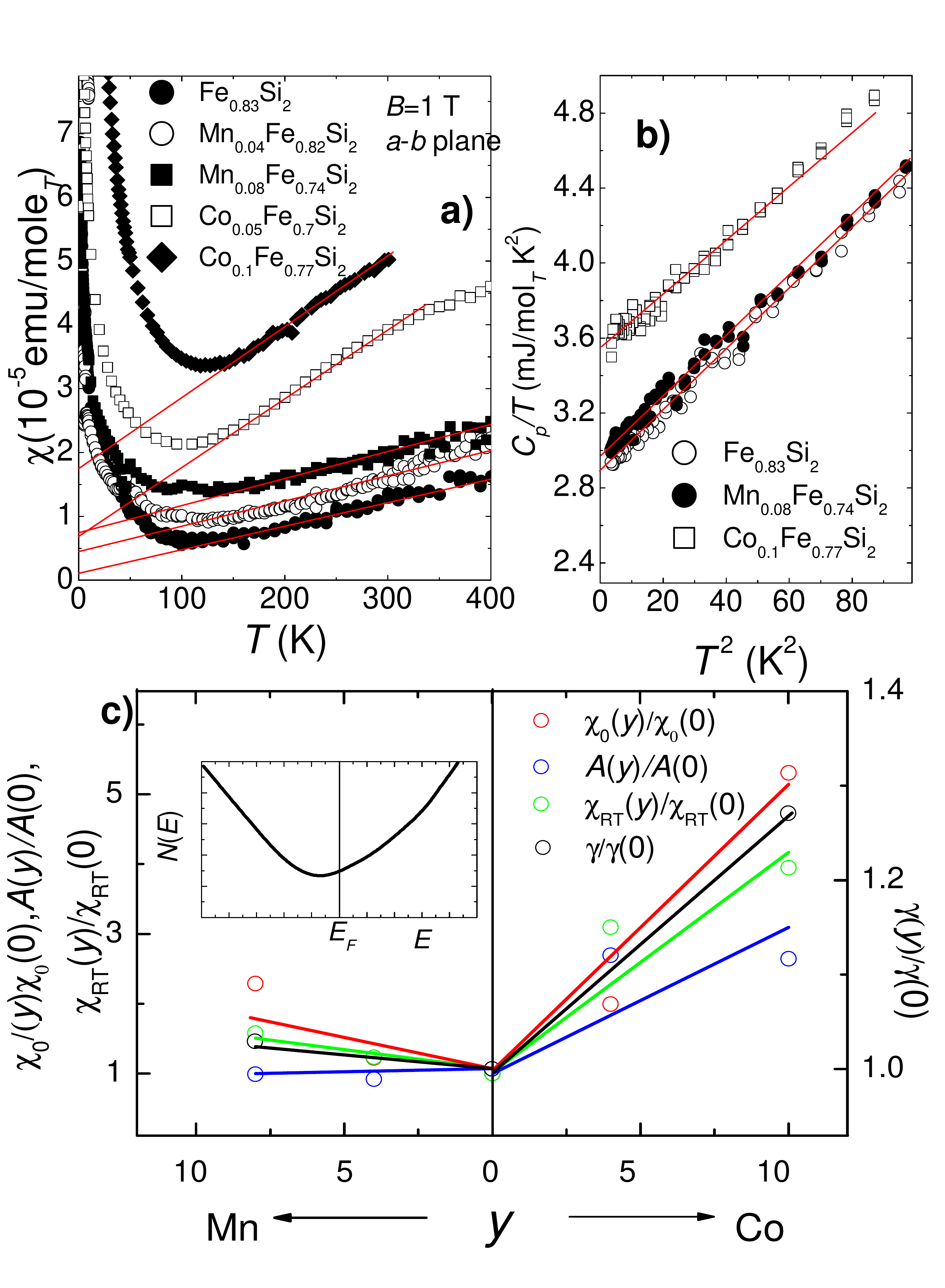}%
\caption{\label{sus_cp_fig}(Color online) a) Magnetic susceptibility of pristine sample of $\alpha-$FeSi$_2$ compared to samples doped with Mn and Co, b) $C_p/T$ ratio vs. $T^2$ for pure, Co and Mn doped $\alpha-$FeSi$_2$ samples. Solid lines are fits to eq. \ref{eq-cpT}. c) Comparison of $\gamma$, $\chi_0$, $\chi_{RT}$ and $A$ yielded for doped Fe$_xT_y$Si$_2$ samples, in relation to the pristine $\alpha$-FeSi$_2$ sample. For details see the text. The inset shows a sketch of the density of states in the vicinity of the Fermi level in undoped $\alpha$-FeSi$_2$.}
\end{figure}

Another way of exploring the electronic structure, especially in the proximity of the Fermi level is chemical doping. Introduction of non-isoelectronic dopants should modify the electronic structure, altering the magnetic and thermodynamic properties of the material. We have doped \afs\ with Co and Mn. Both cations should be incorporated in the structure with nominal 2+ charge, resulting in hole (Mn $d^5$) or electron (Co $d^7$) doping. Four compositions were grown with the general formulae Fe$_xT_y$Si$_2$ -- Mn$_{0.04}$Fe$_{0.74}$Si$_2$, Mn$_{0.08}$Fe$_{0.74}$Si$_2$, Co$_{0.05}$Fe$_{0.7}$Si$_2$ and Co$_{0.1}$Fe$_{0.77}$Si$_2$.
The temperature dependencies of the molar susceptibility $\chi$ for these lightly doped crystals are shown in Fig. \ref{sus_cp_fig}a) together with that of undoped Fe$_{0.83}$Si$_2$. The overall shape of $\chi(T)$ measured for Mn-doped samples resembles that observed for undoped \afs\, in that the slope of $\chi(T)$ above $\simeq$ 100 K is preserved. The only difference is a slight, increase of the susceptibility that corresponds to an additional temperature-independent contribution. The impact of cobalt substitution is more pronounced. The slope of $\chi(T)$ increases drastically, and the overall susceptibility values are higher. A similar effect of Mn and Co doping is apparent in heat capacity measurements (fig. 8b). The Sommerfeld coefficient for Mn doped samples is comparable to that in \afs\ whereas it is strongly enhanced in the Co-doped specimen. 

According to Stoner\cite{stoner1936collective}, the magnetic susceptibility of an itinerant system in a simple, one-band picture should follow $\chi=\chi_0+\alpha T^2$, where $\chi_0$ represents the magnetic susceptibility in the zero temperature limit and $\alpha$ depends on the first and second derivatives of the density of states $N(E)$ with respect to the energy $E$, $N{'}(E)$ and $N{''}(E)$. While this formalism fails to provide a quantitative description of most itinerant paramagnets, it predicts the proper sign of $d\chi {/} dT$. Increasing susceptibility with increasing temperature is expected for systems where the Fermi level is located at the minimum of the $N(E)$. As for all doped samples, as well as \afs\ $d\chi {/} dT<0$ is observed above $\simeq$100 K, and so we expect that the electronic structure at the proximity of the $E_F$ should be consistent with Stoner scenario. Analysis of the magnetic susceptibility, together with measured Sommerfeld ratios $\gamma$, should then give information about the shape of $N(E)$ in proximity to the Fermi level and the $N(E_F)$ itself. For Mn$_{0.08}$Fe$_{0.74}$Si$_2$, the $\gamma$ value is the same as observed for undoped sample (2.9 mJ/mole$_T$), whereas in Co$_{0.10}$Fe$_{0.77}$Si$_2$ is 3.6  mJ/mole$_T$. These values correspond to effective densities of states of 1.25 and 1.5 states/\textit{T} at the Fermi level, respectively.
 
In fig. \ref{sus_cp_fig}c are shown the ratios of $\gamma(y)$, $\chi_{0}(y)$, $\chi_{RT}(y)$ and $A(y)$, i.e. the Sommerfeld coefficient, the intrinsic magnetic susceptibility at the 0 K limit, the room temperature susceptibility and $d \chi \diagup dT$ estimated from the region where $\chi \propto T$ is observed, with their values in the undoped ($y=0$) case. 
The increase of the magnetic susceptibility and the Sommerfeld coefficient upon Co doping is consistent with a rigid band-like shifting of the Fermi level towards a maximum in $N(E)$ at higher energies.
The increase in $A$ reflects a drastic change in the $N(E)$ curvature (inset, Fig. \ref{sus_cp_fig}c). When doping with Mn, on the other hand, $\gamma$, $\chi(T)$ and $A$  change very little, and the rigid-band response of electronic structure fails. Indeed upon shifting the Fermi level towards lower energies, $N(E)$, $N'(E)$, and $N''(E)$ should all be affected. This may happen when the introduced holes are located deep below the $E_F$ or if $dN(E)/d(E)\approx const$. In conclusion we propose that the introduction of holes changes the electronic structure of \afs\ only very little. 

\section{Conclusions}
\label{Conclusions}
Our motivation for this multifaceted investigation, performed for the first time on well-characterized single crystals, was that this intermetallic compound shares a quasi two-dimensional structure with superconducting Fe systems and is a simple Fe-based silicide reference compound to the Fe pnictides. Although there are structural similarities, \afs\ shows a natural tendency to vacancy formation in the transition metal sublattice, together with a lack of electronic correlations. The latter feature is believed to be responsible for magnetism and superconductivity in pnictides and chalcogenides. Lack of electronic correlations is possibly due to the much lower interplanar Fe-Fe distance, and the resulting  more isotropic character of {\afs\}. Broad valence bands and the non magnetic $d^6$ electronic configuration of Fe results in weakly temperature dependent magnetic susceptibility, metallic resistivity, low Sommerfeld coefficient and Wilson ratio close to 1. Theoretical calculations confirm this very weak paramagnetism and DMFT results suggest very weak effective mass enhancement of the charge carriers within the Fe 3$d$ band,  $m^*/m^{LDA} \approx 1.3$. Although our results do not portray \afs\  as a prospective superconductor, we believe that tetralides with square Fe planes may reveal high temperature superconductivity if proper structural and electronic criteria are fulfilled.

Attention should be drawn also to the result of the supercell calculations that imply that the Fe$^{2+}$ valence is independent and protected in the presence of vacancies. We have explained that the linear increase of $\chi(T)$ is a natural result of a dip in density of states at the Fermi level, confirmed by our LDA results. We have examined samples doped with Mn (holes) and Co (electrons), to shed more light on the electronic structure of \afs\ in proximity to the Fermi level. According to magnetic susceptibility and heat capacity data collected for these samples, light doping with Co and Mn does not tune \afs\ towards strong correlations, and even the $N(E_F)$ for heavily ($\simeq $10\%) doped samples is still relatively low. The influence of Co doping may be explained with a rigid-band like shift of the Fermi level within wide minimum at the density of states towards higher $N(E)$, whereas Mn does not change electronic state of \afs\ at all.

\section{Acknowledgements}
We acknowledge the Office of the Assistant Secretary of Defense for Research and Engineering for providing the NSSEFF funds that supported this research.

\bibliography{text_v2}

\begin{thebibliography}{37}%
\makeatletter
\providecommand \@ifxundefined [1]{%
 \@ifx{#1\undefined}
}%
\providecommand \@ifnum [1]{%
 \ifnum #1\expandafter \@firstoftwo
 \else \expandafter \@secondoftwo
 \fi
}%
\providecommand \@ifx [1]{%
 \ifx #1\expandafter \@firstoftwo
 \else \expandafter \@secondoftwo
 \fi
}%
\providecommand \natexlab [1]{#1}%
\providecommand \enquote  [1]{``#1''}%
\providecommand \bibnamefont  [1]{#1}%
\providecommand \bibfnamefont [1]{#1}%
\providecommand \citenamefont [1]{#1}%
\providecommand \href@noop [0]{\@secondoftwo}%
\providecommand \href [0]{\begingroup \@sanitize@url \@href}%
\providecommand \@href[1]{\@@startlink{#1}\@@href}%
\providecommand \@@href[1]{\endgroup#1\@@endlink}%
\providecommand \@sanitize@url [0]{\catcode `\\12\catcode `\$12\catcode
  `\&12\catcode `\#12\catcode `\^12\catcode `\_12\catcode `\%12\relax}%
\providecommand \@@startlink[1]{}%
\providecommand \@@endlink[0]{}%
\providecommand \url  [0]{\begingroup\@sanitize@url \@url }%
\providecommand \@url [1]{\endgroup\@href {#1}{\urlprefix }}%
\providecommand \urlprefix  [0]{URL }%
\providecommand \Eprint [0]{\href }%
\providecommand \doibase [0]{http://dx.doi.org/}%
\providecommand \selectlanguage [0]{\@gobble}%
\providecommand \bibinfo  [0]{\@secondoftwo}%
\providecommand \bibfield  [0]{\@secondoftwo}%
\providecommand \translation [1]{[#1]}%
\providecommand \BibitemOpen [0]{}%
\providecommand \bibitemStop [0]{}%
\providecommand \bibitemNoStop [0]{.\EOS\space}%
\providecommand \EOS [0]{\spacefactor3000\relax}%
\providecommand \BibitemShut  [1]{\csname bibitem#1\endcsname}%
\let\auto@bib@innerbib\@empty
\bibitem [{\citenamefont {Stewart}(2011)}]{RevModPhys.83.1589}%
  \BibitemOpen
  \bibfield  {author} {\bibinfo {author} {\bibfnamefont {G.~R.}\ \bibnamefont
  {Stewart}},\ }\href {\doibase 10.1103/RevModPhys.83.1589} {\bibfield
  {journal} {\bibinfo  {journal} {Rev. Mod. Phys.}\ }\textbf {\bibinfo {volume}
  {83}},\ \bibinfo {pages} {1589} (\bibinfo {year} {2011})}\BibitemShut
  {NoStop}%
\bibitem [{\citenamefont {Tapp}\ \emph {et~al.}(2008)\citenamefont {Tapp},
  \citenamefont {Tang}, \citenamefont {Lv}, \citenamefont {Sasmal},
  \citenamefont {Lorenz}, \citenamefont {Chu},\ and\ \citenamefont
  {Guloy}}]{PhysRevB.78.060505}%
  \BibitemOpen
  \bibfield  {author} {\bibinfo {author} {\bibfnamefont {J.~H.}\ \bibnamefont
  {Tapp}}, \bibinfo {author} {\bibfnamefont {Z.}~\bibnamefont {Tang}}, \bibinfo
  {author} {\bibfnamefont {B.}~\bibnamefont {Lv}}, \bibinfo {author}
  {\bibfnamefont {K.}~\bibnamefont {Sasmal}}, \bibinfo {author} {\bibfnamefont
  {B.}~\bibnamefont {Lorenz}}, \bibinfo {author} {\bibfnamefont {P.~C.~W.}\
  \bibnamefont {Chu}}, \ and\ \bibinfo {author} {\bibfnamefont {A.~M.}\
  \bibnamefont {Guloy}},\ }\href {\doibase 10.1103/PhysRevB.78.060505}
  {\bibfield  {journal} {\bibinfo  {journal} {Phys. Rev. B}\ }\textbf {\bibinfo
  {volume} {78}},\ \bibinfo {pages} {060505} (\bibinfo {year}
  {2008})}\BibitemShut {NoStop}%
\bibitem [{\citenamefont {Taylor}\ \emph {et~al.}(2011)\citenamefont {Taylor},
  \citenamefont {Pitcher}, \citenamefont {Ewings}, \citenamefont {Perring},
  \citenamefont {Clarke},\ and\ \citenamefont
  {Boothroyd}}]{PhysRevB.83.220514}%
  \BibitemOpen
  \bibfield  {author} {\bibinfo {author} {\bibfnamefont {A.~E.}\ \bibnamefont
  {Taylor}}, \bibinfo {author} {\bibfnamefont {M.~J.}\ \bibnamefont {Pitcher}},
  \bibinfo {author} {\bibfnamefont {R.~A.}\ \bibnamefont {Ewings}}, \bibinfo
  {author} {\bibfnamefont {T.~G.}\ \bibnamefont {Perring}}, \bibinfo {author}
  {\bibfnamefont {S.~J.}\ \bibnamefont {Clarke}}, \ and\ \bibinfo {author}
  {\bibfnamefont {A.~T.}\ \bibnamefont {Boothroyd}},\ }\href {\doibase
  10.1103/PhysRevB.83.220514} {\bibfield  {journal} {\bibinfo  {journal} {Phys.
  Rev. B}\ }\textbf {\bibinfo {volume} {83}},\ \bibinfo {pages} {220514}
  (\bibinfo {year} {2011})}\BibitemShut {NoStop}%
\bibitem [{\citenamefont {Zou}\ \emph {et~al.}(2013)\citenamefont {Zou},
  \citenamefont {Feng}, \citenamefont {Logg}, \citenamefont {Chen},
  \citenamefont {Lampronti},\ and\ \citenamefont {Grosche}}]{zou2013fermi}%
  \BibitemOpen
  \bibfield  {author} {\bibinfo {author} {\bibfnamefont {Y.}~\bibnamefont
  {Zou}}, \bibinfo {author} {\bibfnamefont {Z.}~\bibnamefont {Feng}}, \bibinfo
  {author} {\bibfnamefont {P.}~\bibnamefont {Logg}}, \bibinfo {author}
  {\bibfnamefont {J.}~\bibnamefont {Chen}}, \bibinfo {author} {\bibfnamefont
  {G.}~\bibnamefont {Lampronti}}, \ and\ \bibinfo {author} {\bibfnamefont
  {F.}~\bibnamefont {Grosche}},\ }\href@noop {} {\bibfield  {journal} {\bibinfo
   {journal} {arXiv preprint arXiv:1311.0247}\ } (\bibinfo {year}
  {2013})}\BibitemShut {NoStop}%
\bibitem [{\citenamefont {Massalski}\ \emph {et~al.}(1990)\citenamefont
  {Massalski}, \citenamefont {Okamoto}, \citenamefont {Subramanian},\ and\
  \citenamefont {Kacprzak}}]{massalski2001binary}%
  \BibitemOpen
  \bibfield  {author} {\bibinfo {author} {\bibfnamefont {T.~B.}\ \bibnamefont
  {Massalski}}, \bibinfo {author} {\bibfnamefont {H.}~\bibnamefont {Okamoto}},
  \bibinfo {author} {\bibfnamefont {P.~R.}\ \bibnamefont {Subramanian}}, \ and\
  \bibinfo {author} {\bibfnamefont {L.}~\bibnamefont {Kacprzak}},\ }\href@noop
  {} {\emph {\bibinfo {title} {Binary Alloy Phase Diagrams}}}\ (\bibinfo
  {publisher} {ASM International},\ \bibinfo {address} {Materials Park},\
  \bibinfo {year} {1990})\BibitemShut {NoStop}%
\bibitem [{\citenamefont {Birkholz}\ and\ \citenamefont
  {Schelm}(1969)}]{PSSB:PSSB19690340269}%
  \BibitemOpen
  \bibfield  {author} {\bibinfo {author} {\bibfnamefont {U.}~\bibnamefont
  {Birkholz}}\ and\ \bibinfo {author} {\bibfnamefont {J.}~\bibnamefont
  {Schelm}},\ }\href {\doibase 10.1002/pssb.19690340269} {\bibfield  {journal}
  {\bibinfo  {journal} {Physica Status Solidi (b)}\ }\textbf {\bibinfo {volume}
  {34}},\ \bibinfo {pages} {K177} (\bibinfo {year} {1969})}\BibitemShut
  {NoStop}%
\bibitem [{\citenamefont {Dubrovskaya}\ and\ \citenamefont
  {Geld}(1962)}]{dubrovskaya1962structure}%
  \BibitemOpen
  \bibfield  {author} {\bibinfo {author} {\bibfnamefont {L.}~\bibnamefont
  {Dubrovskaya}}\ and\ \bibinfo {author} {\bibfnamefont {P.}~\bibnamefont
  {Geld}},\ }\href@noop {} {\bibfield  {journal} {\bibinfo  {journal} {Russ. J.
  Inorg. Chem.(Engl. Transl.)}\ }\textbf {\bibinfo {volume} {7}},\ \bibinfo
  {pages} {73} (\bibinfo {year} {1962})}\BibitemShut {NoStop}%
\bibitem [{\citenamefont {Gueneau}\ and\ \citenamefont
  {Servant}(1995)}]{gueneau1995quantitative}%
  \BibitemOpen
  \bibfield  {author} {\bibinfo {author} {\bibfnamefont {C.}~\bibnamefont
  {Gueneau}}\ and\ \bibinfo {author} {\bibfnamefont {C.}~\bibnamefont
  {Servant}},\ }\href@noop {} {\bibfield  {journal} {\bibinfo  {journal}
  {Journal of applied crystallography}\ }\textbf {\bibinfo {volume} {28}},\
  \bibinfo {pages} {707} (\bibinfo {year} {1995})}\BibitemShut {NoStop}%
\bibitem [{\citenamefont {Birkholz}\ and\ \citenamefont
  {Frühauf}(1969)}]{PSSB:PSSB19690340270}%
  \BibitemOpen
  \bibfield  {author} {\bibinfo {author} {\bibfnamefont {U.}~\bibnamefont
  {Birkholz}}\ and\ \bibinfo {author} {\bibfnamefont {A.}~\bibnamefont
  {Frühauf}},\ }\href@noop {} {\bibfield  {journal} {\bibinfo  {journal}
  {physica status solidi (b)}\ }\textbf {\bibinfo {volume} {34}},\ \bibinfo
  {pages} {K181} (\bibinfo {year} {1969})}\BibitemShut {NoStop}%
\bibitem [{\citenamefont {Acker}\ \emph {et~al.}(1999)\citenamefont {Acker},
  \citenamefont {Bohmhammel}, \citenamefont {van~den Berg}, \citenamefont {van
  Miltenburg},\ and\ \citenamefont {Kloc}}]{Acker19991523}%
  \BibitemOpen
  \bibfield  {author} {\bibinfo {author} {\bibfnamefont {J.}~\bibnamefont
  {Acker}}, \bibinfo {author} {\bibfnamefont {K.}~\bibnamefont {Bohmhammel}},
  \bibinfo {author} {\bibfnamefont {G.}~\bibnamefont {van~den Berg}}, \bibinfo
  {author} {\bibfnamefont {J.}~\bibnamefont {van Miltenburg}}, \ and\ \bibinfo
  {author} {\bibfnamefont {C.}~\bibnamefont {Kloc}},\ }\href {\doibase
  10.1006/jcht.1999.0550} {\bibfield  {journal} {\bibinfo  {journal} {The
  Journal of Chemical Thermodynamics}\ }\textbf {\bibinfo {volume} {31}},\
  \bibinfo {pages} {1523 } (\bibinfo {year} {1999})}\BibitemShut {NoStop}%
\bibitem [{\citenamefont {Blaauw}\ \emph {et~al.}(1973)\citenamefont {Blaauw},
  \citenamefont {van~der Woude},\ and\ \citenamefont
  {Sawatzky}}]{0022-3719-6-14-017}%
  \BibitemOpen
  \bibfield  {author} {\bibinfo {author} {\bibfnamefont {C.}~\bibnamefont
  {Blaauw}}, \bibinfo {author} {\bibfnamefont {F.}~\bibnamefont {van~der
  Woude}}, \ and\ \bibinfo {author} {\bibfnamefont {G.~A.}\ \bibnamefont
  {Sawatzky}},\ }\href@noop {} {\bibfield  {journal} {\bibinfo  {journal}
  {Journal of Physics C: Solid State Physics}\ }\textbf {\bibinfo {volume}
  {6}},\ \bibinfo {pages} {2371} (\bibinfo {year} {1973})}\BibitemShut
  {NoStop}%
\bibitem [{\citenamefont {Reuther}\ \emph {et~al.}(2001)\citenamefont
  {Reuther}, \citenamefont {Behr},\ and\ \citenamefont
  {Teresiak}}]{0953-8984-13-11-102}%
  \BibitemOpen
  \bibfield  {author} {\bibinfo {author} {\bibfnamefont {H.}~\bibnamefont
  {Reuther}}, \bibinfo {author} {\bibfnamefont {G.}~\bibnamefont {Behr}}, \
  and\ \bibinfo {author} {\bibfnamefont {A.}~\bibnamefont {Teresiak}},\
  }\href@noop {} {\bibfield  {journal} {\bibinfo  {journal} {Journal of
  Physics: Condensed Matter}\ }\textbf {\bibinfo {volume} {13}},\ \bibinfo
  {pages} {L225} (\bibinfo {year} {2001})}\BibitemShut {NoStop}%
\bibitem [{\citenamefont {Aronsson}(1960)}]{aronsson1960note}%
  \BibitemOpen
  \bibfield  {author} {\bibinfo {author} {\bibfnamefont {B.}~\bibnamefont
  {Aronsson}},\ }\href@noop {} {\bibfield  {journal} {\bibinfo  {journal} {Acta
  Chem. Scand}\ }\textbf {\bibinfo {volume} {14}},\ \bibinfo {pages} {1414}
  (\bibinfo {year} {1960})}\BibitemShut {NoStop}%
\bibitem [{\citenamefont {Georges}\ \emph {et~al.}(1996)\citenamefont
  {Georges}, \citenamefont {Kotliar}, \citenamefont {Krauth},\ and\
  \citenamefont {Rozenberg}}]{rmp_dmft}%
  \BibitemOpen
  \bibfield  {author} {\bibinfo {author} {\bibfnamefont {A.}~\bibnamefont
  {Georges}}, \bibinfo {author} {\bibfnamefont {G.}~\bibnamefont {Kotliar}},
  \bibinfo {author} {\bibfnamefont {W.}~\bibnamefont {Krauth}}, \ and\ \bibinfo
  {author} {\bibfnamefont {M.~J.}\ \bibnamefont {Rozenberg}},\ }\href {\doibase
  10.1103/RevModPhys.68.13} {\bibfield  {journal} {\bibinfo  {journal} {Rev.
  Mod. Phys.}\ }\textbf {\bibinfo {volume} {68}},\ \bibinfo {pages} {13}
  (\bibinfo {year} {1996})}\BibitemShut {NoStop}%
\bibitem [{\citenamefont {Pet{\v{r}}{\'\i}{\v{c}}ek}\ \emph
  {et~al.}(2006)\citenamefont {Pet{\v{r}}{\'\i}{\v{c}}ek}, \citenamefont
  {Du{\v{s}}ek},\ and\ \citenamefont {Palatinus}}]{petvrivcek2006jana2006}%
  \BibitemOpen
  \bibfield  {author} {\bibinfo {author} {\bibfnamefont {V.}~\bibnamefont
  {Pet{\v{r}}{\'\i}{\v{c}}ek}}, \bibinfo {author} {\bibfnamefont
  {M.}~\bibnamefont {Du{\v{s}}ek}}, \ and\ \bibinfo {author} {\bibfnamefont
  {L.}~\bibnamefont {Palatinus}},\ }\href@noop {} {\bibfield  {journal}
  {\bibinfo  {journal} {The crystallographic computing system}\ } (\bibinfo
  {year} {2006})}\BibitemShut {NoStop}%
\bibitem [{\citenamefont {Blaha}\ \emph {et~al.}(1990)\citenamefont {Blaha},
  \citenamefont {Schwarz}, \citenamefont {Sorantin},\ and\ \citenamefont
  {Trickey}}]{Blaha1990399}%
  \BibitemOpen
  \bibfield  {author} {\bibinfo {author} {\bibfnamefont {P.}~\bibnamefont
  {Blaha}}, \bibinfo {author} {\bibfnamefont {K.}~\bibnamefont {Schwarz}},
  \bibinfo {author} {\bibfnamefont {P.}~\bibnamefont {Sorantin}}, \ and\
  \bibinfo {author} {\bibfnamefont {S.}~\bibnamefont {Trickey}},\ }\href
  {\doibase 10.1016/0010-4655(90)90187-6} {\bibfield  {journal} {\bibinfo
  {journal} {Computer Physics Communications}\ }\textbf {\bibinfo {volume}
  {59}},\ \bibinfo {pages} {399 } (\bibinfo {year} {1990})}\BibitemShut
  {NoStop}%
\bibitem [{\citenamefont {Haule}\ \emph {et~al.}(2010)\citenamefont {Haule},
  \citenamefont {Yee},\ and\ \citenamefont {Kim}}]{PhysRevB.81.195107}%
  \BibitemOpen
  \bibfield  {author} {\bibinfo {author} {\bibfnamefont {K.}~\bibnamefont
  {Haule}}, \bibinfo {author} {\bibfnamefont {C.-H.}\ \bibnamefont {Yee}}, \
  and\ \bibinfo {author} {\bibfnamefont {K.}~\bibnamefont {Kim}},\ }\href
  {\doibase 10.1103/PhysRevB.81.195107} {\bibfield  {journal} {\bibinfo
  {journal} {Phys. Rev. B}\ }\textbf {\bibinfo {volume} {81}},\ \bibinfo
  {pages} {195107} (\bibinfo {year} {2010})}\BibitemShut {NoStop}%
\bibitem [{\citenamefont {{Yin}}\ \emph {et~al.}(2011)\citenamefont {{Yin}},
  \citenamefont {{Haule}},\ and\ \citenamefont
  {{Kotliar}}}]{2011NatPh...7..294Y}%
  \BibitemOpen
  \bibfield  {author} {\bibinfo {author} {\bibfnamefont {Z.~P.}\ \bibnamefont
  {{Yin}}}, \bibinfo {author} {\bibfnamefont {K.}~\bibnamefont {{Haule}}}, \
  and\ \bibinfo {author} {\bibfnamefont {G.}~\bibnamefont {{Kotliar}}},\ }\href
  {\doibase 10.1038/nphys1923} {\bibfield  {journal} {\bibinfo  {journal}
  {Nature Physics}\ }\textbf {\bibinfo {volume} {7}},\ \bibinfo {pages} {294}
  (\bibinfo {year} {2011})}\BibitemShut {NoStop}%
\bibitem [{\citenamefont {Kutepov}\ \emph {et~al.}(2010)\citenamefont
  {Kutepov}, \citenamefont {Haule}, \citenamefont {Savrasov},\ and\
  \citenamefont {Kotliar}}]{PhysRevB.82.045105}%
  \BibitemOpen
  \bibfield  {author} {\bibinfo {author} {\bibfnamefont {A.}~\bibnamefont
  {Kutepov}}, \bibinfo {author} {\bibfnamefont {K.}~\bibnamefont {Haule}},
  \bibinfo {author} {\bibfnamefont {S.~Y.}\ \bibnamefont {Savrasov}}, \ and\
  \bibinfo {author} {\bibfnamefont {G.}~\bibnamefont {Kotliar}},\ }\href
  {\doibase 10.1103/PhysRevB.82.045105} {\bibfield  {journal} {\bibinfo
  {journal} {Phys. Rev. B}\ }\textbf {\bibinfo {volume} {82}},\ \bibinfo
  {pages} {045105} (\bibinfo {year} {2010})}\BibitemShut {NoStop}%
\bibitem [{\citenamefont {Tomczak}\ \emph {et~al.}(2012)\citenamefont
  {Tomczak}, \citenamefont {Haule},\ and\ \citenamefont {Kotliar}}]{jmt_fesi}%
  \BibitemOpen
  \bibfield  {author} {\bibinfo {author} {\bibfnamefont {J.~M.}\ \bibnamefont
  {Tomczak}}, \bibinfo {author} {\bibfnamefont {K.}~\bibnamefont {Haule}}, \
  and\ \bibinfo {author} {\bibfnamefont {G.}~\bibnamefont {Kotliar}},\
  }\href@noop {} {\bibfield  {journal} {\bibinfo  {journal} {Proc. Natl. Acad.
  Sci. USA}\ }\textbf {\bibinfo {volume} {109}},\ \bibinfo {pages} {3243}
  (\bibinfo {year} {2012})}\BibitemShut {NoStop}%
\bibitem [{\citenamefont {Lee}\ \emph {et~al.}(2008)\citenamefont {Lee},
  \citenamefont {Iyo}, \citenamefont {Eisaki}, \citenamefont {Kito},
  \citenamefont {Fernandez-Diaz}, \citenamefont {Ito}, \citenamefont {Kihou},
  \citenamefont {Matsuhata}, \citenamefont {Braden},\ and\ \citenamefont
  {Yamada}}]{JPSJ.77.083704}%
  \BibitemOpen
  \bibfield  {author} {\bibinfo {author} {\bibfnamefont {C.-H.}\ \bibnamefont
  {Lee}}, \bibinfo {author} {\bibfnamefont {A.}~\bibnamefont {Iyo}}, \bibinfo
  {author} {\bibfnamefont {H.}~\bibnamefont {Eisaki}}, \bibinfo {author}
  {\bibfnamefont {H.}~\bibnamefont {Kito}}, \bibinfo {author} {\bibfnamefont
  {M.~T.}\ \bibnamefont {Fernandez-Diaz}}, \bibinfo {author} {\bibfnamefont
  {T.}~\bibnamefont {Ito}}, \bibinfo {author} {\bibfnamefont {K.}~\bibnamefont
  {Kihou}}, \bibinfo {author} {\bibfnamefont {H.}~\bibnamefont {Matsuhata}},
  \bibinfo {author} {\bibfnamefont {M.}~\bibnamefont {Braden}}, \ and\ \bibinfo
  {author} {\bibfnamefont {K.}~\bibnamefont {Yamada}},\ }\href {\doibase
  10.1143/JPSJ.77.083704} {\bibfield  {journal} {\bibinfo  {journal} {Journal
  of the Physical Society of Japan}\ }\textbf {\bibinfo {volume} {77}},\
  \bibinfo {pages} {083704} (\bibinfo {year} {2008})}\BibitemShut {NoStop}%
\bibitem [{\citenamefont {Wang}\ \emph {et~al.}(2009)\citenamefont {Wang},
  \citenamefont {Wu}, \citenamefont {Wu}, \citenamefont {Liu}, \citenamefont
  {Chen}, \citenamefont {Xie},\ and\ \citenamefont
  {Chen}}]{1367-2630-11-4-045003}%
  \BibitemOpen
  \bibfield  {author} {\bibinfo {author} {\bibfnamefont {X.~F.}\ \bibnamefont
  {Wang}}, \bibinfo {author} {\bibfnamefont {T.}~\bibnamefont {Wu}}, \bibinfo
  {author} {\bibfnamefont {G.}~\bibnamefont {Wu}}, \bibinfo {author}
  {\bibfnamefont {R.~H.}\ \bibnamefont {Liu}}, \bibinfo {author} {\bibfnamefont
  {H.}~\bibnamefont {Chen}}, \bibinfo {author} {\bibfnamefont {Y.~L.}\
  \bibnamefont {Xie}}, \ and\ \bibinfo {author} {\bibfnamefont {X.~H.}\
  \bibnamefont {Chen}},\ }\href
  {http://stacks.iop.org/1367-2630/11/i=4/a=045003} {\bibfield  {journal}
  {\bibinfo  {journal} {New Journal of Physics}\ }\textbf {\bibinfo {volume}
  {11}},\ \bibinfo {pages} {045003} (\bibinfo {year} {2009})}\BibitemShut
  {NoStop}%
\bibitem [{\citenamefont {Kriessman}\ and\ \citenamefont
  {Callen}(1954)}]{PhysRev.94.837}%
  \BibitemOpen
  \bibfield  {author} {\bibinfo {author} {\bibfnamefont {C.~J.}\ \bibnamefont
  {Kriessman}}\ and\ \bibinfo {author} {\bibfnamefont {H.~B.}\ \bibnamefont
  {Callen}},\ }\href {\doibase 10.1103/PhysRev.94.837} {\bibfield  {journal}
  {\bibinfo  {journal} {Phys. Rev.}\ }\textbf {\bibinfo {volume} {94}},\
  \bibinfo {pages} {837} (\bibinfo {year} {1954})}\BibitemShut {NoStop}%
\bibitem [{\citenamefont {{Kojima}}\ \emph {et~al.}(1961)\citenamefont
  {{Kojima}}, \citenamefont {{Tebble}},\ and\ \citenamefont
  {{Williams}}}]{1961RSPSA.260..237K}%
  \BibitemOpen
  \bibfield  {author} {\bibinfo {author} {\bibfnamefont {H.}~\bibnamefont
  {{Kojima}}}, \bibinfo {author} {\bibfnamefont {R.~S.}\ \bibnamefont
  {{Tebble}}}, \ and\ \bibinfo {author} {\bibfnamefont {D.~E.~G.}\ \bibnamefont
  {{Williams}}},\ }\href {\doibase 10.1098/rspa.1961.0030} {\bibfield
  {journal} {\bibinfo  {journal} {Royal Society of London Proceedings Series
  A}\ }\textbf {\bibinfo {volume} {260}},\ \bibinfo {pages} {237} (\bibinfo
  {year} {1961})}\BibitemShut {NoStop}%
\bibitem [{\citenamefont {Stishov}\ \emph {et~al.}(2012)\citenamefont
  {Stishov}, \citenamefont {Petrova}, \citenamefont {Sidorov},\ and\
  \citenamefont {Menzel}}]{PhysRevB.86.064433}%
  \BibitemOpen
  \bibfield  {author} {\bibinfo {author} {\bibfnamefont {S.~M.}\ \bibnamefont
  {Stishov}}, \bibinfo {author} {\bibfnamefont {A.~E.}\ \bibnamefont
  {Petrova}}, \bibinfo {author} {\bibfnamefont {V.~A.}\ \bibnamefont
  {Sidorov}}, \ and\ \bibinfo {author} {\bibfnamefont {D.}~\bibnamefont
  {Menzel}},\ }\href {\doibase 10.1103/PhysRevB.86.064433} {\bibfield
  {journal} {\bibinfo  {journal} {Phys. Rev. B}\ }\textbf {\bibinfo {volume}
  {86}},\ \bibinfo {pages} {064433} (\bibinfo {year} {2012})}\BibitemShut
  {NoStop}%
\bibitem [{\citenamefont {Shimizu}\ and\ \citenamefont
  {Takahashi}(1960)}]{shimizu1960magnetic}%
  \BibitemOpen
  \bibfield  {author} {\bibinfo {author} {\bibfnamefont {M.}~\bibnamefont
  {Shimizu}}\ and\ \bibinfo {author} {\bibfnamefont {T.}~\bibnamefont
  {Takahashi}},\ }\href@noop {} {\bibfield  {journal} {\bibinfo  {journal}
  {Journal of the Physical Society of Japan}\ }\textbf {\bibinfo {volume}
  {15}},\ \bibinfo {pages} {2236} (\bibinfo {year} {1960})}\BibitemShut
  {NoStop}%
\bibitem [{\citenamefont {Mendelsohn}\ \emph {et~al.}(1970)\citenamefont
  {Mendelsohn}, \citenamefont {Biggs},\ and\ \citenamefont
  {Mann}}]{PhysRevA.2.1130}%
  \BibitemOpen
  \bibfield  {author} {\bibinfo {author} {\bibfnamefont {L.~B.}\ \bibnamefont
  {Mendelsohn}}, \bibinfo {author} {\bibfnamefont {F.}~\bibnamefont {Biggs}}, \
  and\ \bibinfo {author} {\bibfnamefont {J.~B.}\ \bibnamefont {Mann}},\ }\href
  {\doibase 10.1103/PhysRevA.2.1130} {\bibfield  {journal} {\bibinfo  {journal}
  {Phys. Rev. A}\ }\textbf {\bibinfo {volume} {2}},\ \bibinfo {pages} {1130}
  (\bibinfo {year} {1970})}\BibitemShut {NoStop}%
\bibitem [{\citenamefont {Lark-Horovitz}\ \emph {et~al.}(1959)\citenamefont
  {Lark-Horovitz}, \citenamefont {edt},\ and\ \citenamefont {Johnson}}]{9120}%
  \BibitemOpen
  \bibfield  {author} {\bibinfo {author} {\bibfnamefont {K.}~\bibnamefont
  {Lark-Horovitz}}, \bibinfo {author} {\bibnamefont {edt}}, \ and\ \bibinfo
  {author} {\bibfnamefont {V.~A.}\ \bibnamefont {Johnson}},\ }\href@noop {}
  {\emph {\bibinfo {title} {Solid state physics :}}}\ (\bibinfo  {publisher}
  {Academic Press,},\ \bibinfo {address} {New York ; London :},\ \bibinfo
  {year} {1959.})\BibitemShut {NoStop}%
\bibitem [{\citenamefont {Junod}\ \emph {et~al.}(1983)\citenamefont {Junod},
  \citenamefont {Jarlborg},\ and\ \citenamefont {Muller}}]{PhysRevB.27.1568}%
  \BibitemOpen
  \bibfield  {author} {\bibinfo {author} {\bibfnamefont {A.}~\bibnamefont
  {Junod}}, \bibinfo {author} {\bibfnamefont {T.}~\bibnamefont {Jarlborg}}, \
  and\ \bibinfo {author} {\bibfnamefont {J.}~\bibnamefont {Muller}},\ }\href
  {\doibase 10.1103/PhysRevB.27.1568} {\bibfield  {journal} {\bibinfo
  {journal} {Phys. Rev. B}\ }\textbf {\bibinfo {volume} {27}},\ \bibinfo
  {pages} {1568} (\bibinfo {year} {1983})}\BibitemShut {NoStop}%
\bibitem [{\citenamefont {Baker}\ \emph {et~al.}(2009)\citenamefont {Baker},
  \citenamefont {Giblin}, \citenamefont {Pratt}, \citenamefont {Liu},
  \citenamefont {Wu}, \citenamefont {Chen}, \citenamefont {Pitcher},
  \citenamefont {Parker}, \citenamefont {Clarke},\ and\ \citenamefont
  {Blundell}}]{1367-2630-11-2-025010}%
  \BibitemOpen
  \bibfield  {author} {\bibinfo {author} {\bibfnamefont {P.~J.}\ \bibnamefont
  {Baker}}, \bibinfo {author} {\bibfnamefont {S.~R.}\ \bibnamefont {Giblin}},
  \bibinfo {author} {\bibfnamefont {F.~L.}\ \bibnamefont {Pratt}}, \bibinfo
  {author} {\bibfnamefont {R.~H.}\ \bibnamefont {Liu}}, \bibinfo {author}
  {\bibfnamefont {G.}~\bibnamefont {Wu}}, \bibinfo {author} {\bibfnamefont
  {X.~H.}\ \bibnamefont {Chen}}, \bibinfo {author} {\bibfnamefont {M.~J.}\
  \bibnamefont {Pitcher}}, \bibinfo {author} {\bibfnamefont {D.~R.}\
  \bibnamefont {Parker}}, \bibinfo {author} {\bibfnamefont {S.~J.}\
  \bibnamefont {Clarke}}, \ and\ \bibinfo {author} {\bibfnamefont {S.~J.}\
  \bibnamefont {Blundell}},\ }\href
  {http://stacks.iop.org/1367-2630/11/i=2/a=025010} {\bibfield  {journal}
  {\bibinfo  {journal} {New Journal of Physics}\ }\textbf {\bibinfo {volume}
  {11}},\ \bibinfo {pages} {025010} (\bibinfo {year} {2009})}\BibitemShut
  {NoStop}%
\bibitem [{\citenamefont {Moroni}\ \emph {et~al.}(1999)\citenamefont {Moroni},
  \citenamefont {Wolf}, \citenamefont {Hafner},\ and\ \citenamefont
  {Podloucky}}]{PhysRevB.59.12860}%
  \BibitemOpen
  \bibfield  {author} {\bibinfo {author} {\bibfnamefont {E.~G.}\ \bibnamefont
  {Moroni}}, \bibinfo {author} {\bibfnamefont {W.}~\bibnamefont {Wolf}},
  \bibinfo {author} {\bibfnamefont {J.}~\bibnamefont {Hafner}}, \ and\ \bibinfo
  {author} {\bibfnamefont {R.}~\bibnamefont {Podloucky}},\ }\href {\doibase
  10.1103/PhysRevB.59.12860} {\bibfield  {journal} {\bibinfo  {journal} {Phys.
  Rev. B}\ }\textbf {\bibinfo {volume} {59}},\ \bibinfo {pages} {12860}
  (\bibinfo {year} {1999})}\BibitemShut {NoStop}%
\bibitem [{\citenamefont {Singh}(2008)}]{PhysRevB.78.094511}%
  \BibitemOpen
  \bibfield  {author} {\bibinfo {author} {\bibfnamefont {D.~J.}\ \bibnamefont
  {Singh}},\ }\href {\doibase 10.1103/PhysRevB.78.094511} {\bibfield  {journal}
  {\bibinfo  {journal} {Phys. Rev. B}\ }\textbf {\bibinfo {volume} {78}},\
  \bibinfo {pages} {094511} (\bibinfo {year} {2008})}\BibitemShut {NoStop}%
\bibitem [{\citenamefont {Nekrasov}\ \emph
  {et~al.}(2008{\natexlab{a}})\citenamefont {Nekrasov}, \citenamefont
  {Pchelkina},\ and\ \citenamefont {Sadovskii}}]{2008JETPL..88..543N}%
  \BibitemOpen
  \bibfield  {author} {\bibinfo {author} {\bibfnamefont {I.}~\bibnamefont
  {Nekrasov}}, \bibinfo {author} {\bibfnamefont {Z.}~\bibnamefont {Pchelkina}},
  \ and\ \bibinfo {author} {\bibfnamefont {M.}~\bibnamefont {Sadovskii}},\
  }\href@noop {} {\bibfield  {journal} {\bibinfo  {journal} {JETP Letters}\
  }\textbf {\bibinfo {volume} {88}},\ \bibinfo {pages} {543} (\bibinfo {year}
  {2008}{\natexlab{a}})}\BibitemShut {NoStop}%
\bibitem [{Note1()}]{Note1}%
  \BibitemOpen
  \bibinfo {note} {The orbital characters refer to the $\protect \sqrt
  {2}\times \protect \sqrt {2}$ non-primitive cell with 2 Fe atoms and space
  group 129, in which the iron planes are oriented as in the conventional cell
  of LiFeAs.}\BibitemShut {Stop}%
\bibitem [{\citenamefont {Nekrasov}\ \emph
  {et~al.}(2008{\natexlab{b}})\citenamefont {Nekrasov}, \citenamefont
  {Pchelkina},\ and\ \citenamefont {Sadovskii}}]{nekrasov2008electronic}%
  \BibitemOpen
  \bibfield  {author} {\bibinfo {author} {\bibfnamefont {I.~A.}\ \bibnamefont
  {Nekrasov}}, \bibinfo {author} {\bibfnamefont {Z.~V.}\ \bibnamefont
  {Pchelkina}}, \ and\ \bibinfo {author} {\bibfnamefont {M.~V.}\ \bibnamefont
  {Sadovskii}},\ }\href@noop {} {\bibfield  {journal} {\bibinfo  {journal}
  {JETP letters}\ }\textbf {\bibinfo {volume} {88}},\ \bibinfo {pages} {543}
  (\bibinfo {year} {2008}{\natexlab{b}})}\BibitemShut {NoStop}%
\bibitem [{Note2()}]{Note2}%
  \BibitemOpen
  \bibinfo {note} {Stripe-like anti-ferromagnetic (AF) order in-plane, and AF
  inter-layer order.}\BibitemShut {Stop}%
\bibitem [{\citenamefont {Stoner}(1936)}]{stoner1936collective}%
  \BibitemOpen
  \bibfield  {author} {\bibinfo {author} {\bibfnamefont {E.~C.}\ \bibnamefont
  {Stoner}},\ }\href@noop {} {\bibfield  {journal} {\bibinfo  {journal}
  {Proceedings of the Royal Society of London. Series A-Mathematical and
  Physical Sciences}\ }\textbf {\bibinfo {volume} {154}},\ \bibinfo {pages}
  {656} (\bibinfo {year} {1936})}\BibitemShut {NoStop}%
\end{thebibliography}%

\end{document}